# Assessing the accuracy of record linkages with Markov chain based Monte Carlo simulation approach

Shovanur Haque[1], Kerrie Mengersen[1] and Steven Stern[1*]
[1]Queensland University of Technology
shovanur.haque@hdr.qut.edu.au, k.mengersen@qut.edu.au, sstern@bond.edu.au

**Abstract**

Record linkage is the process of finding matches and linking records from different data sources so that the linked records belong to the same entity. There is an increasing number of applications of record linkage in statistical, health, government and business organisations to link administrative, survey, population census and other files to create a complete set of information for more complete and comprehensive analysis. To make valid inferences using a linked file, it is increasingly becoming important to assess the linking method. It is also important to find techniques to improve the linking process to achieve higher accuracy. This motivates to develop a method for assessing linking process and help decide which linking method is likely to be more accurate for a linking task. This paper proposes a Markov Chain based Monte Carlo simulation approach, *MaCSim* for assessing a linking method and illustrates the utility of the approach using a realistic synthetic dataset received from the Australian Bureau of Statistics to avoid privacy issues associated with using real personal information. A linking method applied by *MaCSim* is also defined. To assess the defined linking method, correct re-link proportions for each record are calculated using our developed simulation approach. The accuracy is determined for a number of simulated datasets. The analyses indicated promising performance of the proposed method *MaCSim* of the assessment of accuracy of the linkages. The computational aspects of the methodology are also investigated to assess its feasibility for practical use.

---

# 1. Introduction

Record linkage (Newcombe et al. 1959; Fellegi and Sunter 1969) is the process of finding matches and linking records from one or more data sources (e.g., the Census and various health registries or Centrelink datasets) such that the linked records represent the same entity. An entity might be a business, a person, or some other type of listed unit. The term record linkage came originally from the area of public health and also from epidemiological and survey applications (Winkler 1999). In record matching algorithms, records in two files are compared with one another, typically using variables, for example, name, address, and date-of-birth, sex, etc. The individual variables used for connecting records are generally called linking variables or linking fields.

The most commonly used methods in record linkage are deterministic and probabilistic linkage methods. In a deterministic approach, two records are said to be a link if they agree on a high quality identifier (e.g. social security number, tax file number, driver license, etc.) or a combination of identifiers (e.g. first name, date of birth and street name). In a probabilistic method, no unique identifier is available. Record pairs from different files are compared using a set of identifying information comprising one or more linking fields. Each record pair is given a weight based on the likelihood that they are a match. This weight is determined by assessing each linking field for agreement or disagreement, assigning a weight based on this assessment, then summing these individual weights over all linking fields for that pair. This summation is based on the premise of conditional independence, which means that for a record pair the agreement on a linking field is independent of agreement on any other linking field for that pair (Fellegi and Sunter 1969). A decision rule, typically based on a cut-off value, finally determines whether the record pair is asserted to be linked, non-linked or should be considered further as a possible link. Probabilistic record linkage methods are now being well accepted and widely used (Sadinle 2013, 2014, 2016; Steorts 2015, 2016; Herzog et al. 2007; Winkler 2001, 2005).

In recent years, large amounts of data are being collected by organizations in the private and public sectors, as well as by researchers and individuals. Analysing these relevant data can provide huge benefits to businesses and government organizations. Technological advancement now makes it possible to store and process these massive data. However, data from different sources relating to the same entity need to be linked. Moreover, data within a single source may also need to be linked, for example, if there are multiple records for entity over time. Connecting data from different data sources can improve data quality and give better modelling structure (Newcombe et al. 1959, Wallgren 2007; Bakker and Daas 2012). For instance, The Australian Longitudinal Census Database (ACLD) is created by linking the 2006 and 2011 Australian Population Censuses. For the analysis of how characteristics of cohorts change over time, the Australian Bureau of Statistics performed probabilistic linkage of person records in its 2006 and 2011 Census of Population and Housing (Zhang and Campbell 2012). Wilkins et al. (2009) used a linked data set obtained by merging data collected in the Canadian Community Health Survey and data held in Statistics Canada's Hospital Person-Oriented Information database in order to model the relationship between an individual's probability of hospitalization and length of time spent subsequently in hospital and his/her smoking status. In these applications, different data sets related to the same individuals at different points in time are linked.

To make correct inferences using a linked file, it is important to assess the accuracy of the linkages. This motivates two research challenges: to develop a method for assessing the linking process, and to find techniques to improve linking process to achieve higher accuracy where the overall accuracy assessment approach can be used with any method.

Perfect linkage means all records belonging to the same individual are matched and there are no links between records that belong to different individuals. However, in the absence of a unique identifier without error, it is very unlikely to have perfect linkages. This is because linking variables that may be suitable for identifying similar records, such as name, address, date-of-birth etc., may not uniquely identify a person; for example, names may change over time, ages may be entered incorrectly, or addresses may be displayed in different formats, all of which can result in erroneous linkage. In addition to the challenges of missing values, typographical or spelling errors and non-standardized formats of data, sometimes it is hard to identify a correct link even after clerical review. Linkage must also deal with issues of privacy and confidentiality. For example, a person may choose not to enter their age, or individuals' names may not be provided in a de-identified file made available to an analyst or manager.

One way of measuring linkage error is by the proportion of links that are correct matches. Incorrect links create measurement error and bias the analysis (Harron et al. 2014; Chipperfield et al. 2011; Chipperfield and Chambers 2015; Chambers et al. 2009; Lahiri and Larsen 2005). Larsen and Rubin (2001) use the posterior probability of a match for estimating true match status and improve the classification of matches and non-matches through clerical review. However, clerical review can be expensive and time consuming for large databases. Moreover, even after the clerical review it is not possible to be certain about a link being actually correct or incorrect. Lahiri and Larsen (2005) do not consider 1-1 linkage where every record from one file is linked to a distinct record in another file. However, the analytic estimates of precision in Lahiri and Larsen (2005) are poor for 1-1 probabilistic linkage (Chipperfield and Chambers 2015).

As a quality measure, Christen (2012) suggests precision, which is the proportion of links that are true matches. Winglee et al. (2005) use a simulation-based approach, *Simrate* to estimate linkage quality. Their method uses the observed distribution of data in matched and non-matched pairs to generate a large simulated set of record pairs. They assign a match weight to each record pair following specified match rules and use the weight distribution for error estimation. The simulated distribution is used to select an appropriate cut-off for estimating the error rates, but they do not explicitly consider precision. In their simulation approach, they did not simulate the linking process; instead they simulated the comparison outcome for linkage quality measures. Moreover, for the quality measure, most of the work was focused on overall file accuracy.

There has been a lot of work found in the literature which measures the quality of the linked file. However, none of them considered assessing the linking method. In this paper, we are proposing a Markov chain based Monte Carlo simulation approach, *MaCSim* to assess linking method. *MaCSim* is developed to assess a linking method before using the method to link new data files. To achieve this goal, we need two linked files that have been previously linked on similar types of data and link these files to obtain observed links using a defined linking method by *MaCSim*. In this process, the linked files are simulated and relinked using the same linking method. Then the accuracy of these simulated links is calculated by correct relink proportions using observed and simulated links. Based on accuracy results, we can conclude how accurate the linking method is or even whether it is worth linking the new files with this linking method. This approach will also help us to decide which linking method to use to obtain higher accuracy by comparing different linking methods.

The paper is organised as follows. Section 2 describes a problem scenario and a solution idea behind the proposed method. The proposed assessment method, *MaCSim* is described. A range of analyses using the method is described in Section 3. Results of the execution of *MaCSim* on full dataset are provided in Section 4 followed by discussions in Section 5. The paper concludes with a summary and potential future work in Section 6.

## 2. Method

### 2.1 A scenario and solution idea

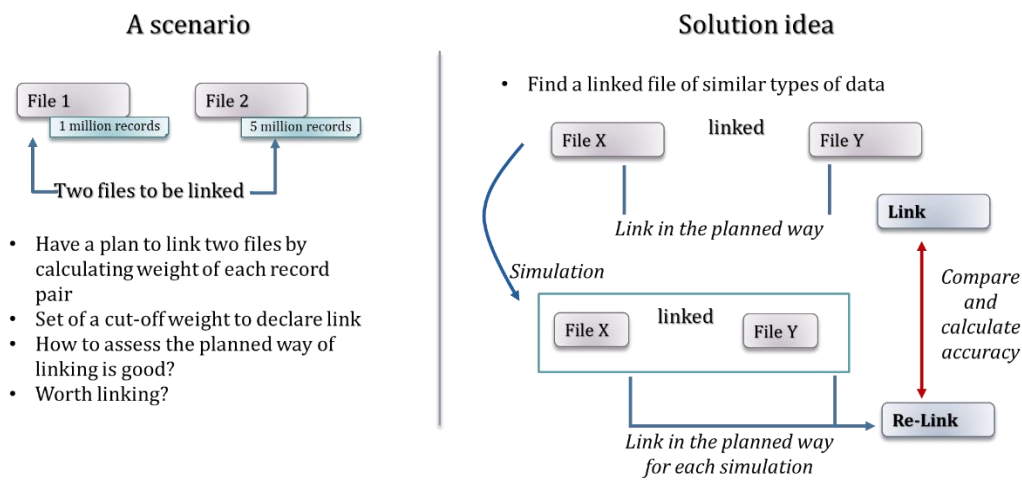

**Figure 1: A problem scenario and possible solution idea**

Figure 1 shows a problem scenario and a solution idea behind the proposed method. When we have two files to link, we need to compare each record pair in these two files to give them a weight and set a cut-off value to declare link. The number of record pairs that need to be compared will be the product of the size of the two files. Thus, for large data files, it will be time consuming. Also, it is hard to know which linking method will give the highest accuracy or even whether it is worth linking these two files.

The goal of the developed methodology is to help on this situation by assessing a linking method using a linked file so that we can decide whether the linking method is useful or not. To achieve this goal, we need two linked files that have been previously linked on similar types of data, link these files using a linking method. We simulate the linked files many times and relink the files using the same linking method. Then we compare these links before and after simulation and calculate accuracy. Based on accuracy results, we can conclude how much accurate the linking method is or even whether it is worth linking these two files.

## 2.2    Proposed solution - *MaCSim*

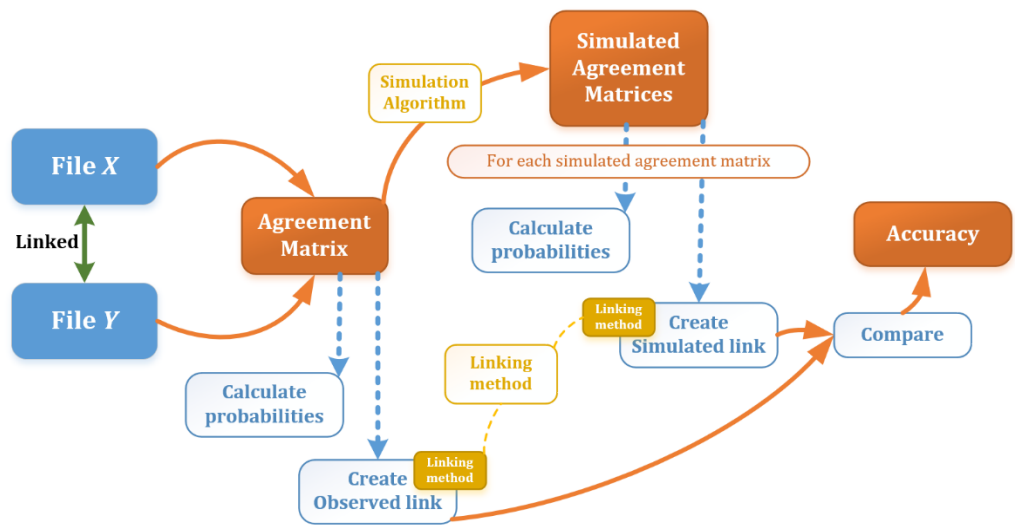

**Figure 2:** *MaCSim*

We proposed a Markov Chain based Monte Carlo simulation method (*MaCSim*) for assessing a linking process or linking method. *MaCSim* utilizes two linked files to create an agreement matrix. From this agreement matrix calculate necessary parameter values and create observed link using a defined linking method. Then, simulate the agreement matrix using a defined algorithm developed for generating re-sampled versions of the agreement matrix. In each simulation with the simulated data, records are re-linked using the same linking method that has been used before simulation. Then the simulated link is compared with the observed link and the accuracy of the individual link is calculated, which ultimately implies the accuracy of the linking method that has been followed to link the records.

*MaCSim* steps involve creating agreement matrix, calculate match and non-match probabilities, create observed link, simulate agreement matrix, calculate probabilities on simulated agreement matrices, create simulated link, calculate accuracy by comparing simulated link with the observed link.

Consider a pair of linked files $X$ and $Y$, where $X$ contains $R_X$ entries and $Y$ contains $R_Y$ entries. There are $L$ linking fields in each file. We define $m_l$ to be the probability that the $lth$ linking field in both files has the same value for a matched pair of records and $u_l$ to be the probability that the $lth$ linking field values in both files are the same for a non-matched pair of records. Further, let $g_l$ be the probability that either or both of the $lth$ linking field values in any record pair are missing regardless of whether the record pair is matched or non-matched. We assume that all missing values occur at random, and denote by $w_l$ the probability that the $lth$ linking field has a value in either file $X$ or file $Y$, individually. Hence, the probability that neither value is missing (from both files) is $1-g_l = (1-w_l)^2$. Therefore, we obtain $w_l = 1 - \sqrt{1 - g_l}$.

## 2.3   Creating agreement matrix *A*

An agreement matrix, **A**, is created from the two files to be linked, $X$ and $Y$, where

$$\boldsymbol{A} = (A_{ijl}); \quad i = 1, \dots R_X, \quad j = 1, \dots R_Y, \quad l = 1, \dots, L,$$

is a three-dimensional array denoting the agreement pattern of all linking fields across all records in the two files. Here, $A_{ijl} = 1$ if the $lth$ linking field value for record $i$ of file $X$ and record $j$ of file $Y$, are the same; $A_{ijl} = -1$ if these values are not the same and $A_{ijl} = 0$ if either or both the values are missing. Therefore, an agreement matrix is a three-dimensional array, contains agreement values 1, -1, and 0, which are the comparison outcome between record pairs of the two files to be linked. We assume that $R_X \leq R_Y$, and each record in file $X$ has a single true matching record in file $Y$. We also assume for simplicity of notation that $A_{iil}$ represents the agreement value of the $lth$ linking field for the true matched record pair in both files.

**Probabilistic record linkage**

The basis of a probabilistic linkage method supposes that there are two files $X$ and $Y$ with records $i$ $and$ $j$, where $i \in X, j \in Y$. All possible pairs of records from these two files can be divided into two disjoint sets $M$ (for matched pair) and $U$ (for non-matched pair). A pair of records will be an element of the set $M$ if they are truly matched (i.e. both represent the same entity). Otherwise, it will be an element of the set $U$ (i.e. represent two different entities). The probabilistic method aims to classify the record pair as an element of either $M$ or $U$. It will be observed whether or not each record pair agrees on the values of the $lth$ linking variable to help decide whether they belong to set $M$ or $U$ (Fellegi and Sunter 1969).

The conditional probabilities $m_l$ and $u_l$ can be written as
$$m_l = Pr\{A_{ijl} = 1 | i,j \text{ a match}\} = Pr\{A_{ijl} = 1 | (i,j) \in M\}$$
$$u_l = Pr\{A_{ijl} = 1 | i,j \text{ not a match}\} = Pr\{A_{ijl} = 1 | (i,j) \in U\}$$

The odds ratio $\frac{Pr\{A_{ijl}|M\}}{Pr\{A_{ijl}|U\}}$ can be used for considering the evidence of $(i,j)$ as a link.

The estimates of $m_l$ and $u_l$ can be used to calculate the odds ratios for agreement and disagreement on $lth$ linking variable. The agreement and disagreement weights are then defined as follows:
$$w_l^{agr} = log_2\left(\frac{m_l}{u_l}\right) if \ record \ pair \ agrees \ on \ linking \ field \ l$$
$$w_l^{dagr} = log_2\left(\frac{1-m_l}{1-u_l}\right) if \ record \ pair \ disagrees \ on \ linking \ field \ l$$

where $w_l^{agr}$ and $w_l^{dagr}$ represent the agreement and disagreement weights for $lth$ linking variable respectively. The base of the logarithm used is immaterial, and base 2 is chosen here as it allows a comparison to information theory results. (Newcombe et al. 1959; Fellegi and Sunter 1969).

## 2.4   Simulating agreement matrix *A*

In order to assess standard errors for estimates deriving from analysis of the linked data, it is of interest to generate re-sampled versions of the agreement matrix **A** in such a way as to preserve the underlying probabilistic linking structure. For this purpose, the *MaCSim* algorithm develops a Markov Chain $\{A^{(n)}\}_{n=0,1,2,...}$ on $A$={set of possible agreement pattern arrays}, with $A^{(0)} = A$, the observed agreement pattern array for the files *X* and *Y*. The key step is to simulate the observed agreement matrix **A** to create **A**\* which includes all the simulated agreement matrices and then apply a linking method to link records using the simulated agreement matrices in each simulation. We estimate the linkage accuracy for each record in every simulation. These estimates are collated and summarized to provide an overall linkage accuracy as described in Section 3.

## 2.5   Simulation algorithm

Markov Chain Monte Carlo (MCMC) (Gelman et al. 1995; Gilks et al. 1996) is an algorithm that constructs a Markov Chain which converges after a certain number of steps to the desired probability distribution and then samples efficiently from this distribution. The generated sample is used as an approximation to the probability distribution for further inference.

The structure of the transition probabilities for the MCMC algorithm employed by *MaCSim* is now outlined. Given the current state of the chain, **A**⁽ⁿ⁾, the next state, **A**⁽ⁿ⁺¹⁾, will be constructed as follows:

*Step 1*: Initially, set $A_{ijl}^{(n+1)} = A_{ijl}^{(n)}$ for all $i, j, and\ l$.

*Step 2*: Randomly select values of $i \in \{1, ..., R_X\}$ and $l \in \{1, ..., L\}$.

*Step 3*: If

a) $A_{iil}^{(n)} = 1$, change $A_{iil}^{(n+1)}$ to –1 with probability $p_1$.

    b) $A_{iil}^{(n)} = -1$, change $A_{iil}^{(n+1)}$ to 1 with probability $p_2$.

*Step 4*: For each $j \neq i$, if

    a) $A_{iil}^{(n)} = 1$ & $A_{iil}^{(n+1)} = -1$, then

        i) If $A_{ijl}^{(n)} = 1$, change $A_{ijl}^{(n+1)}$ to –1.

        ii) If $A_{ijl}^{(n)} = -1$, change $A_{ijl}^{(n+1)}$ to 1 with probability $q_1$.

    b) $A_{iil}^{(n)} = -1$ & $A_{iil}^{(n+1)} = 1$ then

        i) If $A_{ijl}^{(n)} = 1$, change $A_{ijl}^{(n+1)}$ to –1.

        ii) If $A_{ijl}^{(n)} = -1$, change $A_{ijl}^{(n+1)}$ to 1 with probability $q_2$.

    c) $A_{iil}^{(n)} = -1$ & $A_{iil}^{(n+1)} = -1$ then

        If $A_{ijl}^{(n)} = -1$, change $A_{ijl}^{(n+1)}$ to 1 with probability $q_3$.

The values of $p = (p_1, p_2)$ and $q = (q_1, q_2, q_3)$ are described in Section 2.7. Once values for $p$ and $q$ are determined to ensure the stationary distribution of the chain has the desired structure (see Sections 2.6 and 2.7), this Markov chain can be used to generate an appropriate set of re-sampled $\boldsymbol{A}$ values. In particular, we can set $\boldsymbol{A}^{*(s)} = \boldsymbol{A}^{(sd)}$, for $s = 1, \ldots, S$ and some constant $d$. Note that the use of every $d$th member of the chain is designed to reduce the correlation between individual steps which ensures minimal changes in the values of the chain. Also, note that, missingness is static here.

## 2.6    Underlying intuition and maintaining consistency

The transition structure as defined above is designed to replicate circumstances whereby a random element of either file is selected and then a change in its value is made with probability based on its current agreement status with its corresponding partner in the opposite file. Note that if a change does occur, this has the consequent effect of changing the agreement patterns in the associated non-matching record pairs. For instance, if the selected linking variable value in the selected record of the selected file matches its counterpart in the opposite file and was changed, then any agreement indicator for which the associated record in the opposite file was unity (indicating agreement of the values for the selected linking variable) must be re-set to -1, as in steps $4(a)(i)$ and $4(b)(i)$, as they can no longer agree. Alternatively, for non-matched records for which the agreement indicator was -1, the values now may or may not agree, so we reset the indicator value to 1 with the given probability. With this underpinning, it is clear that the internal consistency patterns of agreement will be maintained.

## 2.7 Maintaining marginal distributions

In addition to internal agreement consistency, we need to ensure that the stationary distribution of the Markov chain maintains the required probabilities of agreement for both matched and non-matched records across the two files. This requires appropriate selection of the transition probability parameters $p = (p_1, p_2)$ and $q = (q_1, q_2, q_3)$.

In particular, we require that the probability that linking field values for matched record pairs agree remains equal to $m_l$. That is, $Pr\{A_{iil}^{(n+1)} = 1\} = m_l$.

Assuming that the chain starts in the following state, it is straightforward to see that

$$Pr\{A_{iil}^{(n+1)} = 1\}$$

$$= Pr\{A_{iil}^{(n)} = 1, \text{No Change in Step } 3(a)\} + Pr\{A_{iil}^{(n)} = -1, \text{Change in Step } 3(b)\} =$$
$$Pr\{\text{No Change in Step } 3(a) | A_{iil}^{(n)} = 1\} Pr\{A_{iil}^{(n)} = 1\}$$
$$+ Pr\{\text{Change in Step } 3(b) | A_{iil}^{(n)} = -1\} Pr\{A_{iil}^{(n)} = -1\}$$
$$= (1 - p_1)m_l + p_2(1 - m_l - g_l) = m_l + p_2(1 - m_l - g_l) - p_1 m_l.$$

Thus, we require $p_2 = p_1 m_l/(1 - m_l - g_l)$. Of course, this requirement puts limits on $p_1$, since any value of $p_1 > (1 - m_l - g_l)/m_l$ would result in $p_2 > 1$. However, if $m_l > 0.5(1 - g_l)$ (which it certainly should be for any reasonable and useful linking variable), the necessary constraint of $p_1 < (1 - m_l - g_l)/m_l$ is always satisfied. $p_1$ in this scenario can be thought of as a "mixing rate" parameter and thus the value of $p_1$ should be set as large as possible for using our Markov chain in a computationally efficient manner (i.e. allowing the use of a relatively small value of $d$). This means, without any other constraints, we should select $p_1 = (1 - m_l - g_l)/m_l$ which then implies that $p_2 = 1$. However, as we shall now see whether we can choose this option for $p_1$ depends on the values of $u_l$. In our approach, the key assumption is $(1 - m_l - g_l) \geq 0$, $(1 - u_l - g_l) \geq 0$ and $m_l$, the probability of agreement for matched record pair, should always be greater than the probability of agreement for non-matched record pair, $u_l$ i.e. $m_l > u_l$.

Choosing appropriate values for the $q$ parameters arises from the requirement to maintain the probability of agreement between values of the linking variable among non-matched records. In other words, we must ensure that $Pr\{A_{ijl}^{(n+1)} = 1\} = u_l$. To this end, we note that based on the steps in the algorithm described in Section 2.5, $Pr\{A_{ijl}^{(n+1)} = 1\} = Pr\{A_{ijl}^{(n)} = 1, A_{iil}^{(n)} = -1\} + Pr\{A_{ijl}^{(n)} = 1, A_{iil}^{(n)} = 1,$ No change Step $3(a)\}$

$+Pr\{A_{ijl}^{(n)} = -1, A_{iil}^{(n)} = 1,$ Change Step $3(a)$ & Step $4(a)(ii)\}$

$+Pr\{A_{ijl}^{(n)} = -1, A_{iil}^{(n)} = -1,$ Change Step $3(b)$ & $4(b)(ii)\}$

$+Pr\{A_{ijl}^{(n)} = -1, A_{iil}^{(n)} = -1,$ No change Step $3(b)$ & Change Step $4(c)\}$

$+Pr\{A_{ijl}^{(n)} = 1, A_{iil}^{(n)} = -1,$ No change Step $3(b)\}$

$= u_l w_l + m_l u_l (1 - p_1)/(1 - w_l) + m_l (1 - u_l - g_l) p_1 q_1/(1 - w_l)$
$\qquad + (1 - m_l - g_l)(1 - u_l - g_l) p_2 q_2/(1 - w_l)$
$+(1 - m_l - g_l)(1 - u_l - g_l)(1 - p_2) q_3/(1 - w_l)$
$+(1 - m_l - g_l) u_l (1 - p_2)/(1 - w_l),$

where the above probabilities are calculated based on the relationships between the values of $A_{iil}^{(n)}$ and $A_{ijl}^{(n)}$. For example, $Pr\{A_{ijl}^{(n)} = 0\} = g_l$, but

$Pr\{A_{ijl}^{(n)} = 0 | A_{iil}^{(n)} \neq 0\} = w_l.$

As noted previously, we would like $p_1 = (1 - m_l - g_l)/m_l$ which implies $p_2 = 1$. In this case:

$$Pr\{A_{ijl}^{(n+1)} = 1\} = u_l(2m_l+g_l - 1)/(1 - w_l) + (1 - u_l - g_l)(1 - m_l - g_l)q_1/(1 - w_l) + (1 - m_l - g_l)(1 - u_l - g_l)q_2/(1 - w_l)$$

$$= 1/(1 - w_l)[m_l u_l + (1 - m_l - g_l)\{(-u_l + q_1(1 - u_l - g_l) + q_2(1 - u_l - g_l)\}]$$
$$= 1/(1 - w_l)[m_l u_l + (1 - m_l - g_l)\{(-u_l + (1 - u_l - g_l)(q_1 + q_2)\}]$$

We can readily reduce this to the value of $u_l$ provided $q_1 = q_2 = u_l/(1 - u_l - g_l)$. However, these values are only allowable if $u_l \leq 0.5(1 - g_l)$, as otherwise $u_l/(1 - u_l - g_l)$ exceeds unity.

In the case $u_l > 0.5(1 - g_l)$, we need different values for the $p$ and $q$ parameters. We note that if $u_l > 0.5(1 - g_l)$, then setting $q_1 = q_2 = q_3 = 1$ and

$$p_1 = \frac{(1 - m_l - g_l)(1 - u_l - g_l)}{m_l(3u_l + g_l - 1)}$$

yields $Pr\{A_{ijl}^{(n+1)} = 1\} = u_l$.

Based on the above discussion, in order to maintain the marginal probabilities of matching, we choose the transition probability parameters as follows:

$$p_1 = \begin{cases} (1 - m_l - g_l)/m_l & \text{if } u_l \leq 0.5(1 - g_l) \\ (1 - m_l - g_l)(1 - u_l - g_l)/\{m_l(3u_l + g_l - 1)\} & \text{otherwise} \end{cases}$$

$$p_2 = p_1 m_l/(1 - m_l - g_l)$$

$$q_1 = q_2 = \begin{cases} u_l/(1 - u_l - g_l) & \text{if } u_l \leq 0.5(1 - g_l) \\ 1 & \text{otherwise} \end{cases}$$

$$q_3 = 1.$$

## 2.8 Estimating $m, u$ and $g$ probabilities

In the comparison stage, each linking field value for a record pair from the two files is compared; the result is a ternary code, 1 (when values agree), -1 (when values disagree) and 0 (when either or both values are missing). Hence, the comparison outcomes (i.e. agreement matrix, $A$) contain values 1, -1, and 0. According to these codes, each linking field is given a weight using the probabilities $m$, $u$ and $g$ to recap, $m$ is the probability that the field values agree when the record pair represents the same entity; $u$ is the probability that the field values agree when the record pair represents two different entities, and $g$ is the probability when the field values are missing from either or both records in the pair.

For each linking field using the synthetic data, $m, u, and\ g$ are estimated in the following way:

$m$ = number of values that agree for matched record pairs/total number of matched record pairs.

$u$ = number of values that agree for nonmatched record pairs/total number of nonmatched record pairs.

$g$ = total number of record pairs of which one or both values are missing/total number of possible record pairs.

These probabilities can be estimated using a linked file or they may be known from previous linkages of similar types of data.

## 2.9   Creating observed link

To create the observed links, weights are calculated from the agreement matrix **A** using the probabilities $m, u$ and $g$. For any $(i,j)$-th record pair and any linking variable $l$, if the agreement value is 1 (i.e. $A_{ijl}=1$) then the weight is calculated using $w_{ijl} = log\left(\frac{m_l}{u_l}\right)$; if the value is -1 (i.e. $A_{ijl}=-1$), the weight is calculated using $w_{ijl} = log(1 - m_l - g_l)/(1 - u_l - g_l)$ and for a missing value (i.e. $A_{ijl}=0$), the weight formula is $w_{ijl} = log\ (g_l/g_l) = log\ (1)$.

Given the assumption that missingness occurs at random, and thus has the same chance of occurring in a true matched pair as in a non-match, missing values will not contribute to the weight.

After calculating the weight for each record pair that agree or disagree on a linking variable value explained above, a composite or overall weight, $W_{ij}$ is calculated for each record pair $((i,j)$ by summing individual weights, $w_{ijl}$ over all linking variables for that pair using the following formula:

$$W_{ij} = \sum_{l} w_{ijl}$$

Once weights of all record pairs, $W_{ij}$ are calculated, the observed links are created following the steps of defined linking method below:

 i. First, all record pairs are sorted by their weight, from largest to smallest.
 ii. The first record pair in the ordered list is linked if it has a weight greater than the chosen cut-off value.
 iii. In all the other record pairs that contain either of the records from the associated record pair that have been linked in step b, are removed from the list. Thus, possible duplicate links are discarded.
 iv. Go to step b for the second record and so on until no more records can be linked.

## 3. Results

### 3.1 Data

A synthetic dataset received from the Australian Bureau of Statistics is used for demonstration and analysis to avoid privacy issues associated with using real personal information. Moreover, for synthetic data, it is possible to assign a unique identifier to every record and link them back for verification. Thus, it is possible to calculate the matching quality and validate the accuracy of the model predictions. Many critical issues related to linking process can be investigated by providing controlled conditions with synthetic datasets.

A large file *Y* is generated that comprises 400,000 randomly ordered records corresponding to 400,000 hypothetical individuals. Then, the first 50,000 records are taken to form file *X*. Every record has eight data fields (Table 1). For a record, the value of each variable is generated independently (e.g. the value of BDAY is independent of the value of SA1) and a discrete uniform distribution is used to generate its value except the value of COB. 300,000 records are assigned a value '1101' for 'Born in Australia'. The remaining 100,000 records are randomly assigned one of about 300 country codes according to the corresponding proportion of people in the 2006 Australian Census. In file *X*, the RECID (Record Identifier) stays matched to the *Y* file for each record. This makes it easy to identify true matches and non-matches in the linking process.

**Table 1: Data field description**

| Data field | Value |
|---|---|
| RECID (Record identifier) | 7 alphanumeric characters ranges from 'A000001' to 'A400000'. |
| SA1 (Statistical Area 1) | a hypothetical two-level geographical location system, Statistical Area 1 (SA1). Each SA1 contains exactly 400 records. The values are 5 digit code numbered from 10001 to 11000. |
| MB (Meshblock) | Every SA1 consists of exactly 5 Meshblocks or MB. Each Meshblock contains 80 records of file $Y$ and 10 records in file $X$. The values are 7 digit code ranges from 1000101 to 1100009. |
| BDAY (Birth Day) | 20,000 consecutive days from 1 January 1955 to 3 October 2009. BDAY values are numeric and ranges from 1 to 366. |
| BYEAR (Birth Year) | Value is numeric and ranges from 1955 to 2009. |
| SEX (Male/Female) | The value 1 and 2 represents male and female respectively. Exactly 50% of all records are male, and the rest 50% are female. |
| EYE (Eye Colour) | Values are numbered from 1 to 5 and are evenly distributed. |
| COB (Country of Birth) | 75% of the total records are assigned a value '1101' for 'Born in Australia'. The remaining 25% records are randomly assigned one of about 300 country codes according to the corresponding proportion of people in the 2006 Census. |

Some values in file $X$ are changed intentionally to simulate errors in linking fields. The value of a variable in file $X$ is changed by replacing it either with a randomly chosen value from the records in file $Y$ or setting the value to 'missing'. For this modification, individual records are selected independently. The SA1 field is changed to an adjacent SA1 for 500 (1%) records, and the first five digits of the corresponding Meshblock code are altered appropriately. For 1,500 (3%) records, the MB is changed to another MB within the same SA1 region. BDAY is changed to 'missing' for 4,000 (8%) records. For 500 records (1%), the day and month corresponding to the numeric code are altered. In the BYEAR field, 50 records are replaced with 'BYEAR–2', 50 with 'BYEAR+2'. 1200 records are reset to 'BYEAR–1' and 1200 to 'BYEAR+1'. For the SEX field, the value of 50 records (0.1%) is reversed. For 5,000 records (10%), the value of EYE field is set to 'missing'. For another 5,000 records (10%) a valid alternative is chosen as a replacement value. The COB field is set to 'missing' for 750 records (approximately 2%) of the records coded to "1101". COB is also set to 'missing' for 250 records (approximately 2%) with another country code. For 125 of these cases, records are replaced with 'Australia' and for the remaining 125 cases, records in COB are recoded to another country within the same broad geographical region (e.g. with the same two-digit SACC code) (Peter Rossiter 2014).

### 3.2  Blocking strategy

In the linking process, the number of possible record pairs to compare will depend on the size of the two files. For large data files, comparing and calculating weights of each record pairs can cause a significant performance bottleneck. Moreover, it is not computationally efficient and often not possible to undertake matching algorithms which search through entire large data files to find matches. To overcome these challenges, the files are split into blocks where the matches are most likely. Thus, blocking reduces the large number of comparisons by only comparing record pairs that have the same value for a blocking variable.

The analysis used two different blocking variables, namely SA1, and SA1 & SEX. We combine two variables as a blocking variable. For every blocking variable, the number of records in each block in file $X$ is different. Due to the introduced misclassification error described above, the values of the variable SA1 are changed in file $X$. Therefore, while blocking with SA1, we took the original value of SA1 to make sure all the true matches are within this block. Similarly, while blocking with SA1 & SEX, we consider the original values of this combined variable. For the analysis, seven variables (i.e. SA1, MB, BDAY, BYEAR, SEX, EYE, COB) are used. Following the specific blocking strategy, an agreement array, $A$ is created from the two files to be linked for a single block. Block-specific $m, u \text{ and } g$ probabilities are calculated for each linking variable, following the procedure described in Section 2.8.

### 3.3  Simulation - creating simulated values of $A$

The initial agreement matrix $A$ is simulated following the steps described in Section 2.5. The thinning value $d$ is set as 1,000 and the number of desired replicates of $A$, say $A^*$, is $S$ = 1,000. Hence, 1,000,000 MCMC simulations are run and $s$ samples $A^{(s)}, s = 1, \ldots, 1000$, are retained. In $A^*$, we have 1000 instances of the agreement matrix $A$. The "thinning" parameter allows us to specify whether and how much the MCMC chains should be thinned out in order to reduce the correlation between the elements of the MCMC sample. In our case, a thinning value $d$=1000 results in keeping every 1000th value and discarding all interim values.

## 3.4 Examine simple distance between $A^*$ entries

The distances between $A^*$ entries ($A^{*(2)}$, $A^{*(3)}$, ...., $A^{*(S)}$) from the initial agreement matrix $A^{*(1)}$ is calculated. In every simulation, the distance is calculated by the total number of agreement values that are changed from the initial values divided by the total number of agreement values. In this way we obtain the proportion of agreement values that are changing in each simulation.

Figure 3 shows the distances in 1000 simulations using: (i) blocking variable SA1, (ii) combined variable SA1 & SEX.

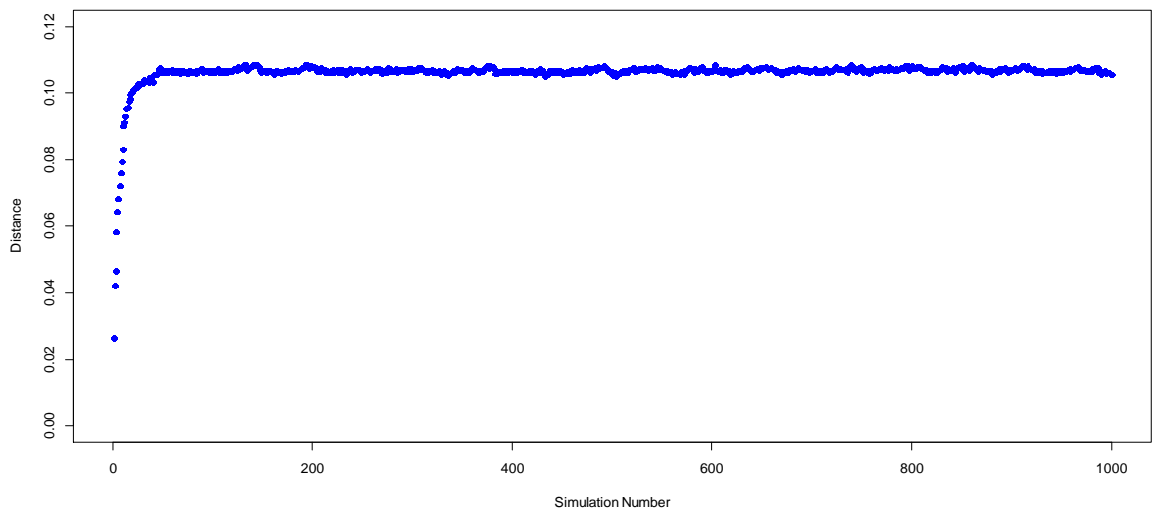

**(i) blocking variable SA1**

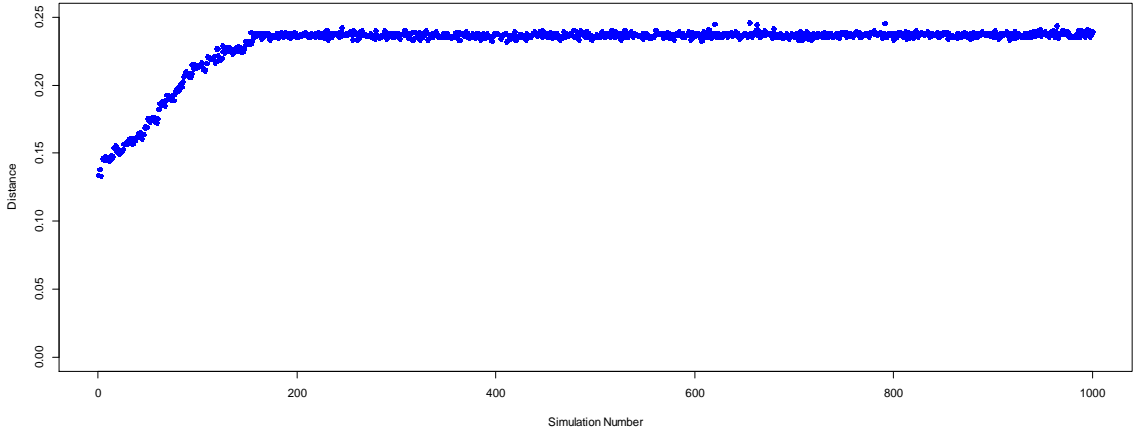

(ii) combined variable SA1 and SEX

**Figure 3: Distance of $A^*$ entries from the initial agreement matrix for blocking variable (i) SA1 and (ii) Combined SA1 and SEX**

For SA1, the distance plot allows estimation of a "burn-in" period for the chain as well as the thinning parameter ($d$) to ensure that the retained simulated matrices are less correlated. From the distance graph on blocking variable 'SA1' (Figure 3(i)), the chain appears to have converged after 50 iterations when approximately 11% of the values in the elements of $A^*$ are changed. The chain stays stable in 1000 simulations. In the case of the combined blocking variable SA1 & SEX, we see from the plot (Figure 3 (ii)) that the chain converges after 180 iterations to around 0.24. Hence compared to the single blocking variable SA1, the chain for the combined variable took more iteration to settle in.

Table 2 shows the comparison of the percentage of agree, disagree, and missing values for blocking variable SA1 and SA1_SEX. From the table we noticed that the percentage of agree is higher in case of SA1 compared to SA1_SEX. Since in the simulation algorithm, the changes of agreement/disagreement values in the next state depends on the agreement/disagreement values of the current state; thus, for these two variables the total number of values changes in each simulation is expected to be different. Therefore, the convergence occurs in two different points for these two blocking variables.

Table 2: Percentage of Agree, Disagree and missing for each blocking variable

|  | Block size | All record pairs | | | Matched record pairs | | | Non-matched record pairs | | |
| --- | --- | --- | --- | --- | --- | --- | --- | --- | --- | --- |
|  |  | "1" | "-1" | "0" | "1" | "-1" | "0" | "1" | "-1" | "0" |
| SA1 | 59 | 24.1% | 72.5% | 3.4% | 92.9% | 3.1% | 4.0% | 23.9% | 72.7% | 3.4% |
| SA1_SEX | 26 | 19.2% | 77.8% | 3.0% | 92.6% | 3.7% | 3.7% | 18.8% | 78.2% | 3.0% |

## 3.5 Total number of agreement values changes in $A^*$ in each simulation

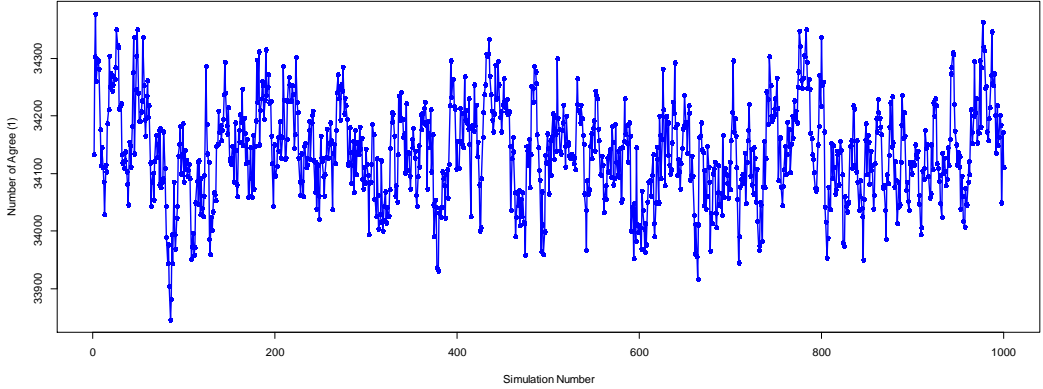

**(i) Total number of agree (1) in $A^*$**

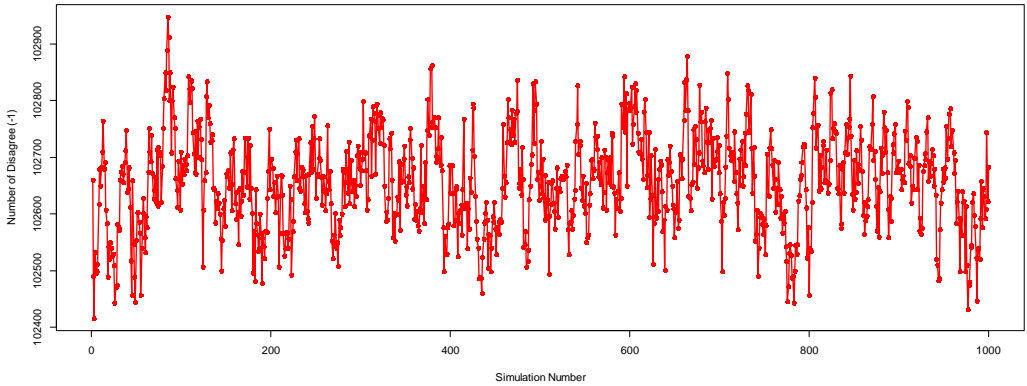

**(ii) Total number of disagree (-1) in $A^*$**

**Figure 4: Total number of agreement values,**

**(i) agree, 1 and (ii) disagree, -1 in $A^*$ in each simulation**

The agreement matrices in $A^*$ contain ternary values, 1 for agree, -1 for disagree and 0 for missing when we compare a record pair for each variable. In every simulation these three agreement values are changed following our defined algorithm (Section 2.5). Figure 4 (i) shows the total number of agree (1) values in each simulation among all the 141,600 (=59x400x6, with 59 records in file $X$, 400 records in file $Y$ and 6 linking variables) agreement values in each agreement matrix inside $A^*$. Similarly, Figure 4 (ii) shows the total number of disagree (-1) values in each simulation among the 141,600 agreement values in each agreement matrix inside $A^*$. The difference in total number of agree (1) and disagree (-1) values in each simulation indicates the changes made by the algorithm. The missing (0) values are not shown as these are kept static and do not contribute to the weight.

### 3.6 Agreement value changes in $A^*$ for a record pair in each simulation

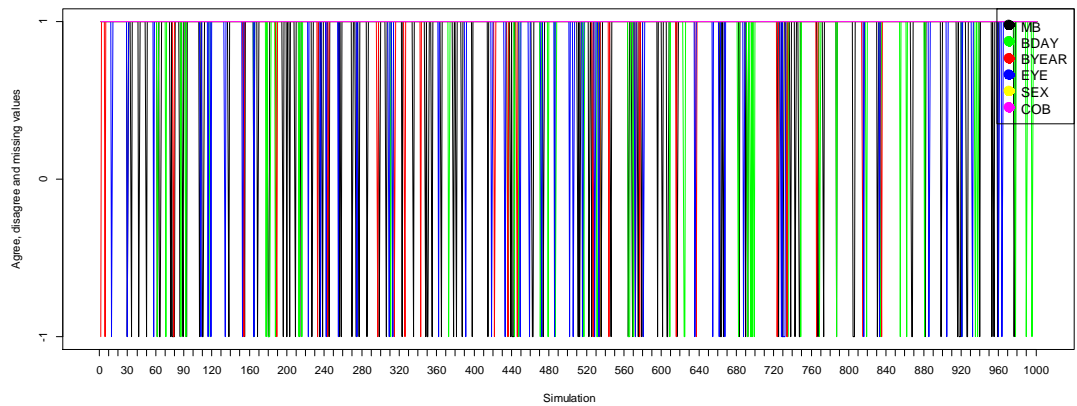

**Figure 5: Agreement values (agree, 1; disagree, -1; missing, 0) changes in $A^*$ for a record pair for every linking variable**

Figure 5 shows the changes in the agreement values (1, -1, and 0) of six linking variables for one record pair in each of the 1000 retained simulations in $A^*$. Here, each coloured line represents each linking variable values and the distribution of these lines over 1000 simulations proves the changes of agreement values made by the algorithm from one simulation to the next.

### 3.7 Proportion of times each record in File *X* is correctly re-linked

Based on the agreement values from $A^*$, in every simulation we link records following the same linking method described earlier (Section 2.9) and observe how many times each record has been re-linked to the record to which it was originally linked. We perform this analysis on the first block when blocking with SA1 and also with combined variable SA1 & SEX.

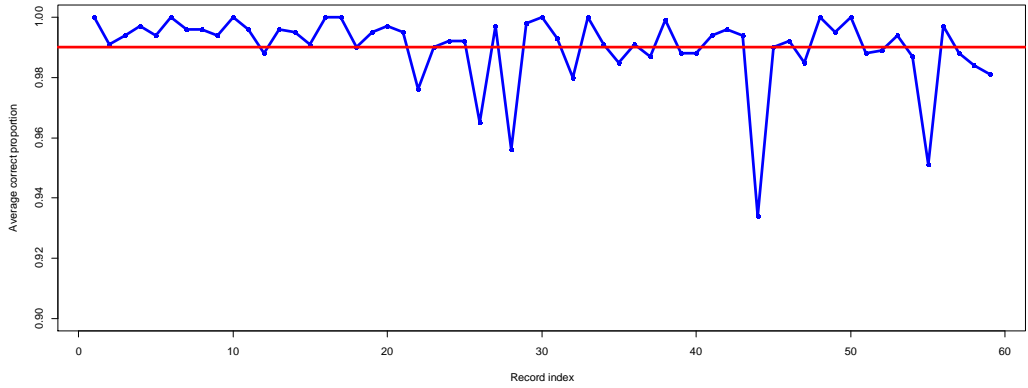

**(i) Correct re-link proportion (first block of SA1)**

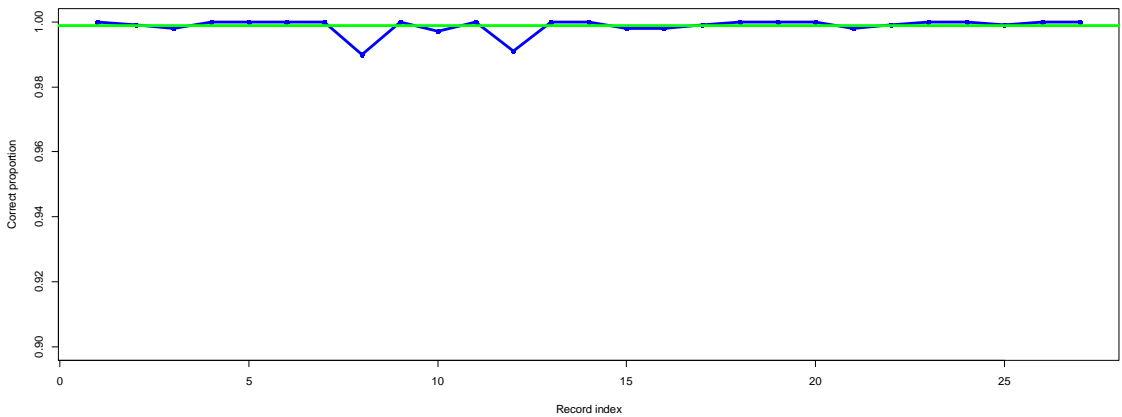

**(ii) Correct re-link proportion (first block of SA1 & SEX)**

**Figure 6: Correct re-link proportion of each *X* record**

When we block the data with SA1, the first block contains 59 records in file *X*. Figure 6 (i) shows the proportion of correct links of each *X* record for this block in 1000 simulations. From this plot, we see that the correct re-link proportion for all 59 records lies between 93.5% and 100%. The plot also shows the average accuracy with the red line, which is 99%. We have a very low error rate for each record. The maximum error we obtained was 6.5% for record number 44.

With the combined variable SA1 & SEX (Figure 6 (ii)), there are 26 records in the first block in file *X*. Figure 6 (ii) shows the correct re-link proportion of each *X* record in 1000 simulations. Here we obtained an accuracy in excess of 98%. The average accuracy is 99.8% which is shown by the green line. The maximum error is only 1.2%, for record number 8.

## 3.8 Correct re-link proportion in every simulation

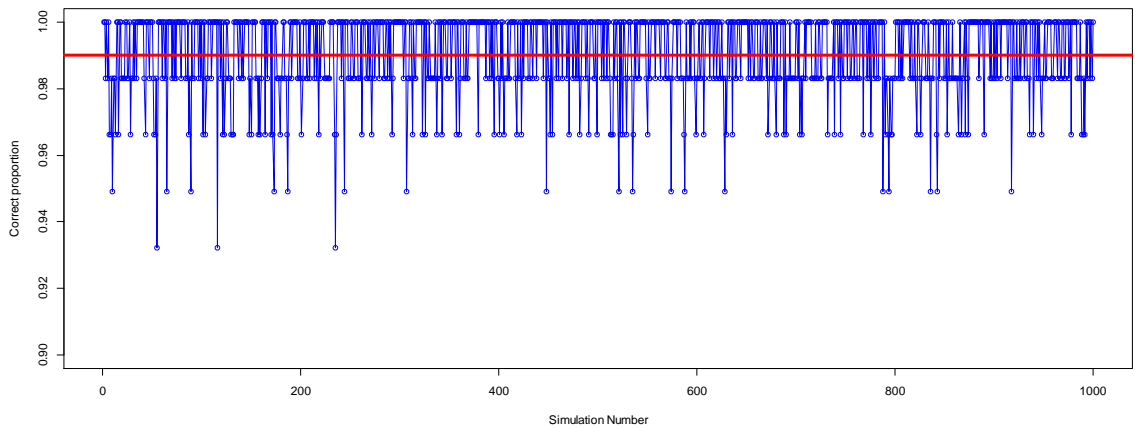

**(i) Correct re-link proportion in every simulation (first block of SA1)**

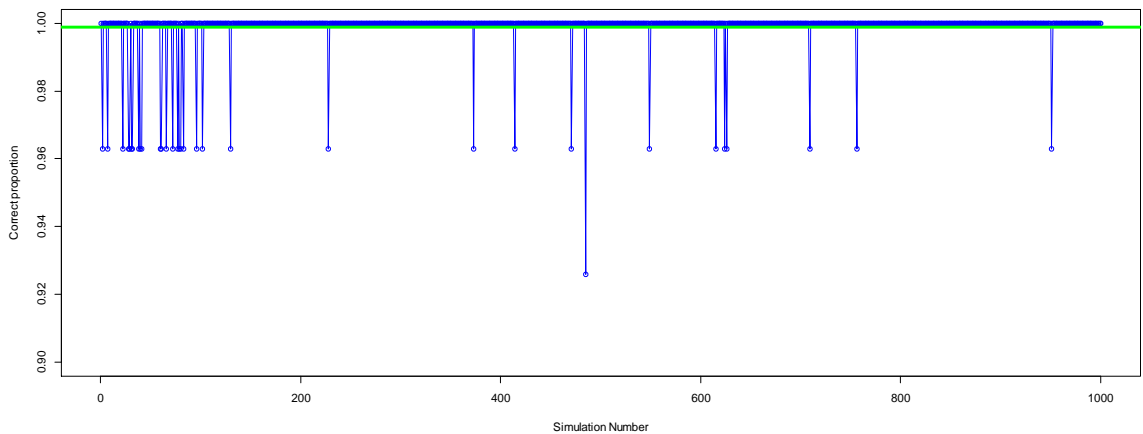

**(ii) Correct re-link proportion in every simulation (first block of SA1 & SEX)**

**Figure 7: Correct re-link proportion in every simulation**

In this analysis, we estimate the accuracy in every simulation for all records in File *X* for the first block when blocking with variable SA1 and also with the combined blocking variable SA1 & SEX. The plot (Figure 7 (i)) shows the correct re-link proportion of all 59 records in each of 1000 simulations. We obtained 100% accuracy in most of the simulations. For some simulations 98.3% accuracy is obtained where 58 records (out of 59) are correctly linked to the original records and one record is incorrectly re-linked. The smallest accuracy, 93.2% (=55/59), is found in only three simulations where 4 records are incorrectly linked. Note that the average accuracy (indicated with the red line in Figure 7 (i)) for all records in every simulation is 99%, which is exactly the same as the average accuracy for each record in all simulations (Figure 6 (i)).

With the combined blocking variable SA1 & SEX Figure 7 (ii) shows the correct re-link proportion of all 26 records in each of 1000 simulations. We obtained 100% accuracy in most of the simulations. For some simulations 96.1% accuracy is obtained where 25 records (out of 26) are correctly linked to its original records. The smallest accuracy, 92.3% (=24/26), is found in only one simulation where 2 records are incorrectly linked. Note that the average accuracy (indicated with the green line in Figure 7 (ii)) for all records in every simulation is 99.8%, which is exactly the same as the average accuracy for each record in all simulations (Figure 6 (ii)).

## 4. *MaCSim* on full dataset

We execute *MaCSim* on full dataset. With 50,000 records in file *X* and 400,000 records in file *Y* with 1,000 blocks and with 6 variables, a total of 120 millions comparisons are made in each simulation. We performed 1 million simulations using *MaCSim* algorithm and retained every 1000$^{th}$ simulated agreement matrix to create overall simulated outcome. Blocking strategy reduced the overall simulation execution time. For a dataset, it is possible to save the overall simulated outcome and reuse it for analysis instead of going through the simulation process each time. However, relinking based on simulated agreement matrices took a rather long time for calculating weights for all record pairs in each simulation.

**Execution setup:**

The code ran on an IBM Thinkpad with the processor of Intel i7 and memory of 16GB. *MaCSim* is executed in each block one after another and the results for all blocks are concatenated.

In the table (Table 3) we showed the execution time of each step of *MaCSim* on full dataset. Total execution time is 227.5 hours. Although, the execution time looks reasonable, various optimization techniques, such as parallelisation can improve the execution time.

**Table 3: Full dataset execution time:**

| MaCSim steps | Each block (sec) |
|---|---|
| Create agreement matrix | 4 |
| Calculate block-specific probabilities | < 1 |
| Create observed links | 1 |
| Simulate agreement matrix to create A* | 83 |
| Calculate probabilities from simulated agreement matrices | 35 |
| Create simulated links | 695 |
| Calculate proportion of correct links | 1 |
| **Total execution times** | **819** |

**Table 4: Correct re-link proportion result on full dataset:**

| Correct re-link proportions | Number of records | Percentage | Correct re-link proportions | Number of records | Percentage |
|---|---|---|---|---|---|
| 1.00 to 0.90 | 49,529 | 99.06% | 1.00 to 0.99 | 38,765 | 77.53% |
| 0.90 to 0.80 | 416 | 0.83% | 0.99 to 0.98 | 6,330 | 12.66% |
| 0.80 to 0.70 | 44 | 0.09% | 0.98 to 0.97 | 2,015 | 4.03% |
| 0.70 to 0.60 | 5 | 0.01% | 0.97 to 0.96 | 892 | 1.78% |
| 0.60 to 0.50 | 7 | 0.01% | 0.96 to 0.95 | 457 | 0.91% |
| 0.50 to 0.40 | 0 | 0.00% | 0.95 to 0.94 | 338 | 0.68% |
| 0.40 to 0.30 | 4 | 0.008% | 0.94 to 0.93 | 242 | 0.48% |
| 0.30 to 0.20 | 1 | 0.002% | 0.93 to 0.92 | 205 | 0.41% |
| 0.20 to 0.10 | 0 | 0.00% | 0.92 to 0.91 | 147 | 0.29% |
| 0.10 to 0.00 | 0 | 0.00% | 0.91 to 0.90 | 132 | 0.26% |

We calculated correct re-link proportion for each of 50,000 records in every simulation. We found more than 99% records have correct re-link proportion more than 90% (Table 4). The right side of the table shows (more granular results) correct re-link proportions of records between 90% to 100%. We found 90% records have correct re-link proportions more than 98%. Figure 8 shows the average correct re-link proportion of each $X$ record in 1000 simulations. From this plot, we see that the average proportion for all 50,000 records is 99.1% indicated by the red line. These proportions are also shown by the density in Figure 9 and histogram in Figure 10.

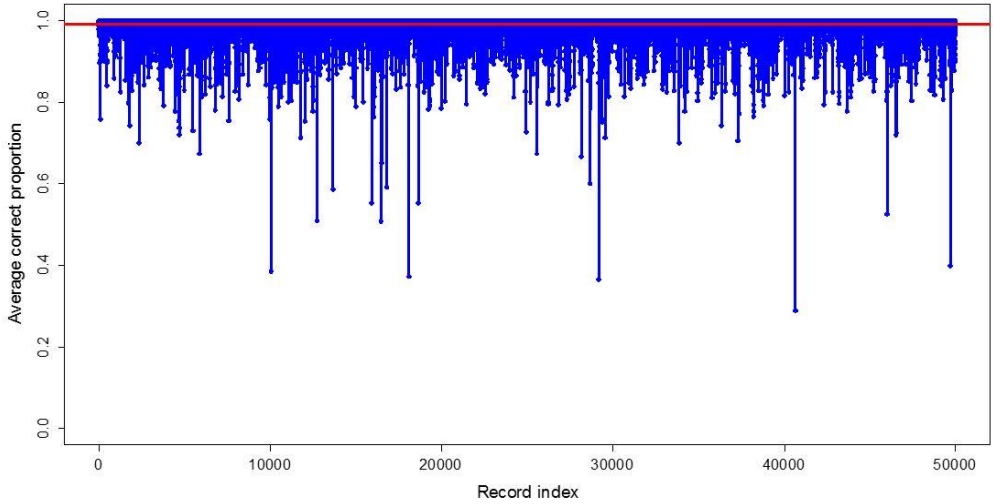

**Figure 8: Correct re-link proportion of each record – average is 0.9911 (red line)**

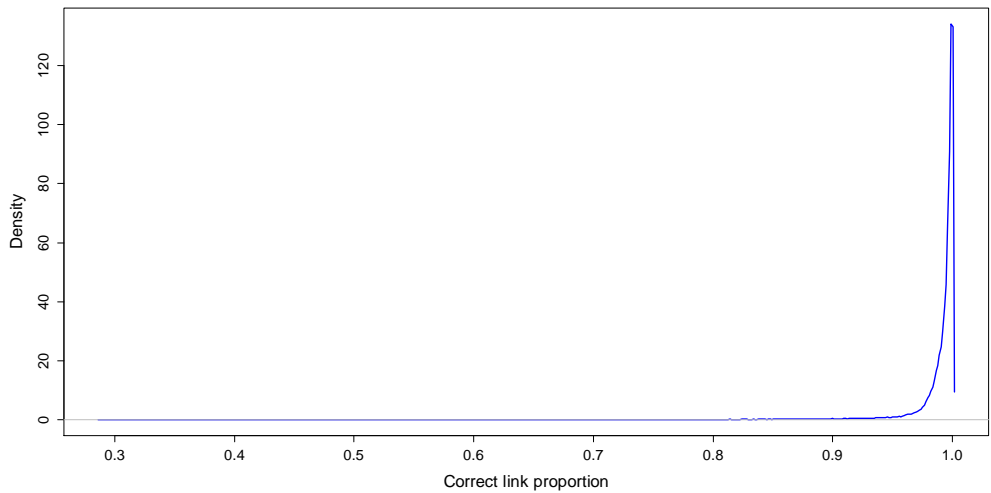

**Figure 9: Correct re-link proportion of each record – Density plot**

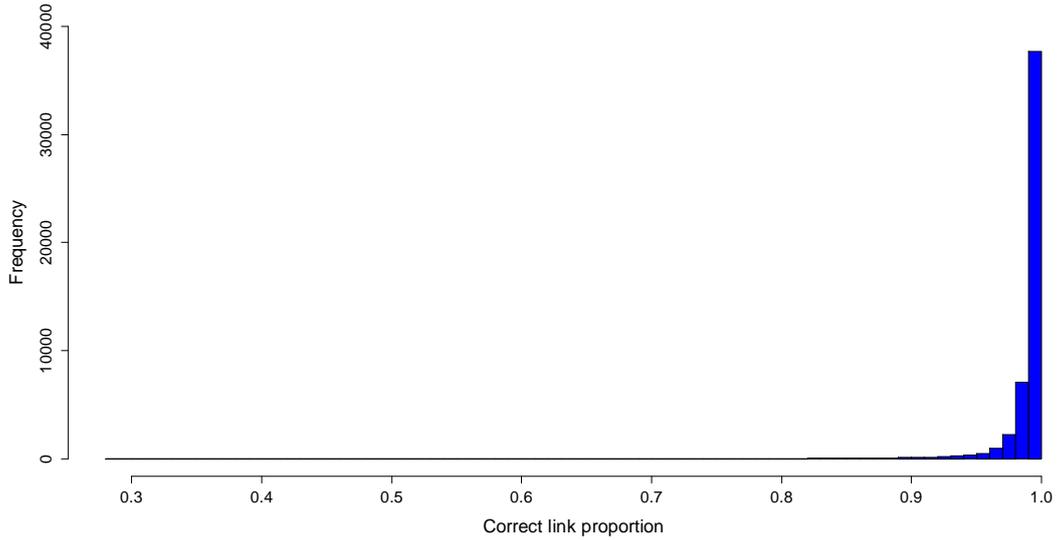

**Figure 10: Correct re-link proportion of each record – Histogram**

## 5. Discussions

When there is a task to link two files, it is hard to decide which method to use for linking. Since these are new files, there is no way to measure the accuracy after linking without further review. *MaCSim* can assist in the evaluation of which method will give higher accuracy to link these files. *MaCSim* needs two linked files that have been previously linked on similar types of data to ultimately assess or help decide which linking method to use for linking new files or even whether it is worth linking the files. Match and non- match probabilities can be estimated using a linked file or they may be known from previous linkages of similar types of data. The approach can be used as a tool to assess a linking method, or to evaluate or compare other linking methods. Based on the obtained accuracy results, the user can decide on a preferred method or evaluate whether it is worth linking the two files at all.

The *MaCSim* algorithm develops a Markov Chain $\{A^{(n)}\}_{n=0,1,2,\ldots}$ on $A$={set of possible agreement pattern arrays}, with $A^{(0)} = A$, the observed agreement pattern array for the files $X$ and $Y$. The structure of the transition probabilities for the chain is outlined. Once values for these probabilities are determined to ensure the stationary distribution of the chain has the desired structure, this Markov chain can be used to generate an appropriate set of re-sampled $A$ values. It is shown that in addition to internal agreement consistency, our chain maintains the required probabilities of agreement for both matched and non-matched records across the two files. Therefore, transition probability parameter values are derived to maintain the marginal probabilities of matching.

The Markov chain employed by *MaCSim* can be used to generate an appropriate set of re-sampled **A** (agreement matrix) values. In our study, 1,000,000 MCMC simulations are run and $s$ samples $A^{(s)}$, $s = 1,....,1000$, are retained. In $A^*$, we have 1000 instances of the agreement matrix **A**. Note that the use of every **d**th member (thinning parameter $d = 1000$) of the chain is designed to reduce the correlation between individual steps which results in minimal changes in the values of the chain. The characteristics of $A^*$ is observed (Section 3.5 & 3.6) by looking at the changes of agreement values (agree, disagree, and missing) made by the algorithm from one simulation to the next.

In MCMC sampling, once the chain has converged, its elements can be seen as a sample from the target posterior distribution. The distance plots (Figure 3) for both single and combined variables show convergence of the chain. By the nature of the MCMC process, the elements of the sample can be highly correlated. The parameter "thinning" allows us to specify whether and by how much the MCMC chains should be thinned out in order to reduce this correlation. In our case, a thinning value **d**=1000 results in keeping every 1000th value and discarding all interim values. This ensures that the retained simulated matrices are less correlated.

*MaCSim* measures the average accuracy of each record in all simulations and average accuracy for all records in every simulation (Figure 6 & 7). In both cases, we obtained average accuracy, which is 99%. Therefore, the linking method used in *MaCSim* could be a better choice to link new files of similar types of data as it gave high accuracy. Alternatively, the user can test other methods using *MaCSim* and compare the accuracy results to decide which method to choose to link the new files.

Two different blocking variables, namely SA1 (Statistical Area 1), and combined variable SA1 & SEX are used for the analyses. Other variables e.g., MB, BYEAR, BDAY can also be used for blocking and testing. When we blocked with these variables, we had only a small number of records to link (for MB, there were 6 records in one block) and the correct re-link proportions for those records exceeded 95%. Considering the purpose of the proposed method, we elected to present results with only a couple of blocking variables.

Probabilistic linkage is widely used in the absence of unique identifier. Our *MaCSim* approach is tested on numeric data for the case study. The method compares records, checks for similarity and assigns values (1, -1, and 0) according to the match and non-match between records. *MaCSim* can also be used on text data fields, in which case data needs to be prepared, such as by exploiting text similarity functions, the processes of parsing, standardisation etc.

*MaCSim* has been implemented in R (programming language) and the computational aspects of the methodology is investigated. The code is stable, parameterized and reusable on different sets of data. Blocking is used to reduce computational time. The computational time can be further reduced by using high performance computing (HPC) and applying optimization techniques, such as parallelisation.

The *MaCSim* approach is tested on two datasets. The method is yet to be investigated on multiple datasets. In *MaCSim*, missingness is considered static and the effect of missing data patterns on the accuracy of record linkage is ignored. *MaCSim* also did not consider the effect of conditional independence assumptions (which means for a record pair, the agreement on a linking field is independent of agreement on any other linking field for that pair) on linkage accuracy.

## 6. Conclusion

With ever expanding overlapping datasets, both administrative and substantive, the need to accurately assess the linkage of these databases is crucial. It is also important for assessing which, if any, linking method is likely to be more accurate for a linkage task. This paper proposed a Markov Chain based Monte Carlo simulation approach *MaCSim* which can be used as a tool to assess a linking method, or to evaluate or compare other linking methods. Based on the obtained accuracy results, the user can decide on a preferred method or evaluate whether it is worth linking at all. The accuracy is determined for a number of simulated datasets, and therefore will better represent uncertainty than an estimate from just one dataset. The linking method that is used in *MaCSim* could be a better choice to perform linking as it gave promising results.

*MaCSim* approach is tested on numeric data for the case study. The method can also be used on text data fields, in which case data needs to be prepared, such as by exploiting text similarity functions, the processes of parsing, standardisation etc.

We have implemented *MaCSim* in R (programming language) and investigate the computational aspects of the methodology. Test results show robust performance of the proposed method of assessment of accuracy of the linkages. Blocking is used to reduce computational time. The computational time can be further reduced by using high performance computing (HPC) and applying optimization techniques, such as parallelisation. Furthermore, for a dataset, it is possible to save the overall simulated outcome and reuse it for analysis instead of going through the simulation process each time. The simulated agreement matrix can be stored and reuse to assess other linking methods.

Future work is to investigate the approach to assess the effect of missing information and conditional independence assumptions on linkage accuracy and enhance the Markov chain methodology to account for the case of conditional dependence. Moreover, in this work we have used two datasets; how the approach will work on more than two datasets is yet to be investigated.